\documentclass[onecolumn,aps,prl]{revtex4}
\usepackage{graphicx}
\usepackage{epstopdf}
[1999/03/29 PSNFSS v.7.2
Times font as default roman
: S Rahtz]

\begin{document}
\renewcommand{\thefootnote}{\fnsymbol{footnote}}
\sloppy
\newcommand{\rp}{\right)}
\newcommand{\lp}{\left(}
\newcommand \be  {\begin{equation}}
\newcommand \ba {\begin{eqnarray}}
\newcommand \bas {\begin{eqnarray*}}
\newcommand \ee  {\end{equation}}
\newcommand \ea {\end{eqnarray}}
\newcommand \eas {\end{eqnarray*}}

\title{Pricing stocks with yardsticks and sentiments}
\thispagestyle{empty}

\author{Sebast\'ian Mart\'inez Bustos$^1$, J\o rgen Vitting Andersen$^2$, 
Michel Miniconi$^3$, 
Andrzej Nowak$^4$,  
Magdalena Roszczynska-Kurasinska$^4$ and David Br\'ee$^4$  
\vspace{0.5cm}}
\affiliation{$^1$Department of Mathematics, Unversidad de los Andes, 
Bogot\'a, Colombia
\\}
\affiliation{$^2$CNRS, Institut Non Lin\'eaire de Nice
1361 route des Lucioles, Sophia Antipolis 
F06560 Valbonne, France\\} 
\affiliation{$^3$Laboratoire J.-A. Dieudonn\'e Universit\'e de Nice-Sophia 
Antipolis, Parc Valrose 06108 Nice Cedex 02, France\\}
\affiliation{$^4$Department of Psychology, Warsaw University, 00-183 Warsaw, Poland\\}
\email{vitting@unice.fr}


\date{\today}

\begin{abstract}
{\bf 
Human decision making by professionals trading daily in the stock market can be a daunting task.
It includes decisions on 
whether to keep on investing or to exit a market subject to huge price swings, and 
how to price in news or rumors attributed to a specific stock. 
The question 
then arises how professional traders, who specialize in daily 
buying and selling large amounts of a given stock, know 
how to properly price a given stock on a given day? 
Here 
we introduce the idea that people use heuristics, or ``rules of thumb'', in terms of ``yard sticks'' 
from the performance of the other stocks in a stock index. The under-/over-performance 
with respect to such a yard stick then signifies a general negative/positive sentiment of 
the market participants towards a given stock. 
Using empirical data of  
 the Dow Jones Industrial Average, stocks are shown to have daily performances 
with a clear tendency to cluster around the measures introduced by the yard sticks. 
We illustrate how sentiments, most likely due to insider 
information, can influence the performance of a given stock over period of months, 
and in one case years.  
 }
\end{abstract}

\maketitle

\vspace{1cm}

\section{Introduction} 
One of the founders of Behavioral Finance, D. Kahneman\cite{Shefrin},
 once pointed out the close resemblance in the media coverage of 
financial markets to a stereotypical individual. As he mentioned, the media often describe  
financial markets 
with attributes like  ``... thoughts, beliefs, moods and sometimes stormy emotions. 
The main characteristic of the market is extreme nervousness. It is full of hope 
one moment and full of anxiety the next day ...''. 
One way to get a first quantification  of 
the sentiment of the market is to probe the sentiments of its investors. 
Studying sentiments of consumers/investors 
 and its impact on markets has become an increasingly important topic. A various number of sentiment 
indices of investors/consumers already exists, some of which have now been recorded 
in a time span over some few decades. 
The Michigan Consumer 
Sentiment index, published monthly by 
the University of Michigan and Thomson Reuteurs, is probably the one which has the largest direct 
impact on markets when published. The natural question then arises as to whether it is possible 
to predict market movements based on the sentiments of consumers/investors? 

Fisher and Statman \cite{FS} made a study of tactical asset allocation from data of the sentiment 
of a heterogeneous group (large,medium,small) 
 of investors. 
The main idea in \cite{FS} was to look for indicators for future stock returns 
based on the diversity of sentiments. The study found the sentiments of different groups 
do not move in lockstep, and that sentiments for the groups of large and small investors, 
could be used as contrary indicators for future 
S\&P 500 returns. Similar results were found by Baker and Wurgler who showed that 
when beginning-of-period proxies for sentiments are low, subsequent returns are 
relatively high for securities whose valuations are highly subjective and difficult to 
arbitrage\cite{BW}
However other papers are reporting on feedback loops between sentiment and market performance: 
past market returns determine investors sentiment who often expect a continuation of short 
term returns\cite{BC}. 
Recent research\cite{Tetlock} on investors sentiment expressed in the media (as 
measured from the daily content of a {\em Wall Street Journal} column) seem to point in this 
direction with high media pessimism predicting downward pressure on market prices. Such results are 
 in line with theoretical models of noise and liquidity traders\cite{DSSW1,DSSW2}. Other studies\cite{AF} 
claim very little predictability on stock returns using computational linguistics to extract sentiments on 
 1.5 million internet message boards posted on {\em Yahoo! Finance} and {\em Raging Bull}. However in that 
study it was shown that disagreement induces trading and message posting activity was shown to 
correlate with volatility of the market. 

Common to all the aforementioned studies is the aim to 
predict {\bf global} market movements from sentiments either obtained in surveys or internet message 
boards. In the following we instead propose to 
obtain a sentiment related pricing 
of a given asset by expressing the {\bf relative} sentiment of a given stock to the market. This is  
similar to the principle of the Capital Asset Price Model\cite{CAPM} 
(CAPM) which relates the price of a given asset to 
that of the price of the market, instead of trying to give absolute the price level of the asset/market. 
Put differently: we will in the following introduce a method that does not estimate the impact of 
a given ``absolute'' level of sentiment can have on the market, but instead introduce a sentiment 
of an asset {\em relative} to the sentiment of the general market whatever the absolute (positive/negative) 
sentiment of the general market. As will be illustrated this gives rise to a pricing formula for a given stock 
relative to the general market, much like the CAPM, but now with the relative sentiment of the stock to 
the market included in the pricing.

\noindent
\section{Theory}
We will in the following consider how traders find the appropriate price level 
of a given stock on a daily basis. One could for example have in mind traders that specialize on a given 
stock, who follow actively its price movements, maybe to consider opportune moments to 
either buy some amounts or instead sell as a part of a larger order. 
The question is what influences the decision making for traders when to enter and when 
to exit the market? According to the standard economic view only expectations 
about future earning/dividends 
and future interest rate levels should matter in the pricing of a given stock. Looking at the often 
very big price swings during earnings or interest rate 
 announcements, this part clearly seem to play a major role 
at least at some specific instances of time. But what about other moments when there is no news which can 
be said to be relevant for future earnings/interest rates? The fluctuations seen in daily 
stock prices simply can not be explained by new information related 
to these two factors, nor can risk aversion,  so why do stock prices 
fluctuate so much and how do traders navigate the often rough seas of fluctuations? 

Here we will take a heuristics point of view, and argue that traders need some rule of thumb, 
or as we prefer to refer to it, ``yardsticks'', in order to know how to position themselves.  
A first rough estimate for a trader would obviously be the returns of other stocks 
in a given stock index. Let us denote $s_i(t)$ the daily (nominal) return of stock $i$ belonging to 
the given index at 
time $t$, 
and $R_{- i}(t)$ the return of the remaining 
$N-1$ stocks in the index at time $t$. We emphasis to {\em exclude}  the contribution 
of stock $i$ in $R_{-i}$ in 
order to avoid any self-impact which would amount to assuming that ``the price of a stock 
rises because it rises''.
Using the averaged return of the other stocks as a first crude 
yardstick one would then have that traders of stock $i$ would price the stock according 
to 
\be 
\label{s_1}
s_i(t) \simeq 
 R_{- i}(t) \equiv  {1 \over (N-1)} \sum_{j \neq i} s_j(t)
\ee

A powerful tool to check an equation, often used in Physics, is to use dimensional analysis and 
ensure that the quantities of both sides have the same dimensions. By the same token, an expression should 
be independent of the unit used. Since eq.~(\ref{s_1}) expresses a relationship between 
returns, i.e. a quantity that expresses increments in percentages, 
it is already dimensionless. However we argue that there is a mental
relevant ``unit'' in play, 
which is the size of a typical daily fluctuation of a given stock\cite{JVA}. Such a mental ``unit 
of fluctuation'' is created by 
the memory of traders who follow closely past performance of a given stock. Since the capacity of 
human working memory is quite small, investors are not able to analyze all available data. 
This causes individuals to simplify the world around them\cite{BEA}.  Investors need to choose consciously 
or unconsciously data that seem to be the most informative. While doing it (whatever the mode they are in) 
investors are exposed to the 
recency effect according to which they will remember recent prices better than earlier ones\cite{MC}. 
The most recent prices are stored in short-term memory and therefore are easier to be retrieved  than 
earlier ones that are stored in long-term  memory. Hereby, the last volatility of the stock  becomes 
a ``unit of fluctuations'' which investors refer to while anticipating future price changes. This 
``unit'' which enters their attention field, influences their ability to estimate 
probabilities in accordance with the rule that events that are easier to be retrieved 
from the memory seem to be more likely to people than they really are\cite{TK}.
Dividing (\ref{s_1}) on each side by the size
of a typical fluctuation would therefore be one way to ensure independence of such  
units. Taking the standard deviation as measure of a typical fluctuation, 
 the renormalized (\ref{s_1}) takes the 
form:
\be
\label{s_2}
{ s_i(t) \over  
 \sigma(s_i) } = 
 {R_{- i}(t)
 \over \sigma(R_{- i}) } .
\ee
Here the standard deviation $\sigma (X) $ of the variable $X$ 
is defined over a given window size $T$ from 
the variance $ \sigma^2 = {\rm Var}  \equiv E_T(X^2) -  E_T^2(X)$, with $E$ denoting the 
expectation value.

As we will show in a moment, for most stocks over daily time periods of 
time (\ref{s_2}) turns out to be a good 
approximation. There are however strong and persistent deviations. Actually we will in 
the following define stocks for which   
 (\ref{s_2}) holds on average, as ``neutral'' 
with respect to the sentiment of the traders. Similarly we 
 use the relation 
as a measure of how biased (positive or negative) a sentiment traders have on 
the given 
stock. More precisely the sentiment of a given stock $i, \alpha_i$, is defined as:
\be
\label{alpha}
\alpha_i(t) = 
 {s_i(t) \over \sigma (s_i) }
- 
 {R_{- i}(t)
 \over \sigma (R_{- i}) } . 
\ee
We emphasize that the sentiment is defined with respect to the other stocks in the index, 
which serves as the neutral reference. 
The ratio of a stock's (excess) return to its standard deviation, tells something about its 
performance, or reward-to-variability ratio, also called the Sharpe ratio in 
Finance\cite{Sharpe}.
Therefore (\ref{alpha}) attributes a positive (respective negative) bias/sentiment to a stock, 
$\alpha_i >0$ (respective $\alpha_i <0$), when the 
Sharpe ratio of  
 the stock exceeds (respective underperforms) the Sharpe ratio of the sum of the other stocks in 
the index\cite{note2}.

Rewriting (\ref{alpha}) the pricing of stock $i$ can now be given in  
terms of a renormalized performance of the other stocks in the index as well as an 
eventual  bias:
\be
\label{s_3}
 s_i(t) = 
 \sigma (s_i) 
\alpha_i(t)  
+ 
 {\sigma (s_i)  
 \over \sigma (R_{- i}) } 
 R_{- i}(t)
\ee
As a first check of (\ref{s_3}) we take the expectation 
value of (\ref{s_3}) by averaging over all stocks that an index is composed of, 
and average over time (daily returns). A priori,
over long periods of time one would expect 
to find as many positive biased as negative biased stocks in an 
 index composed of many stocks. That this is indeed the case will be shown empirically. 
Using this 
assumption 
the term with $\alpha_i$ 
disappears due to symmetry. One gets: 
\be
\label{our_hyp}
 E(s_i) = 
{ E(R_{-i}) 
 \over \sigma (R_{-i})  }
 \sigma (s_i)
\ee

\noindent(\ref{our_hyp}) is very similar in structure 
to the famous Capital Asset Pricing Model in Finance (CAPM)\cite{note1}: 
\ba
\label{capm}
{E(s_i) - R_f \over \beta_i} & = &
 E(R) - R_f ; \beta_i = {{\rm Cov}(s_i,R) \over \sigma^2 (R)}  
\ea
$R_f$ in (\ref{capm}) is the risk free return which, since we consider daily returns, will 
be taken equal 0 in the following:
\ba
\label{capm1}
E(s_i) & = &
{{\rm Cov} (s_i,R) \over \sigma^2 (R)} 
  E(R) . 
\ea
Apart from the exclusion of the stock itself in the expression $R_{-i}$, 
the main difference between the CAPM in the form (\ref{capm1}) and our hypothesis 
(\ref{our_hyp}) is that we 
stress the use of standard deviations in the pricing formula  
 instead of the covariance between the stock return 
and the index return on the right side of (\ref{capm1}). 
Furthermore we argue that the covariance between a given stock and the index 
is not a very stable measure over time, whereas this is the case for the variance of 
a given stock. One reason for instability of the covariance 
could for example be created by sudden ``shocks'' in 
terms of specific bad or good news for a given company. After such a ``shock'' 
we postulate that the covariance between the stock and the index changes, whereas 
the stocks variance remain the same but with a change in relative 
performance as expressed through 
 (\ref{alpha}). (\ref{our_hyp}) is reminiscent of the so-called capital allocation line 
in Finance which expresses the return of a portfolio composed of a certain 
percentage of the market portfolio and the remaining invested in a risk free asset. 
The capital allocation line however only expresses the return of this specific portfolio, 
whereas our expression is supposed to hold for each individual asset.

\section{Results}
\begin{figure}[h]
\includegraphics[width=14cm]{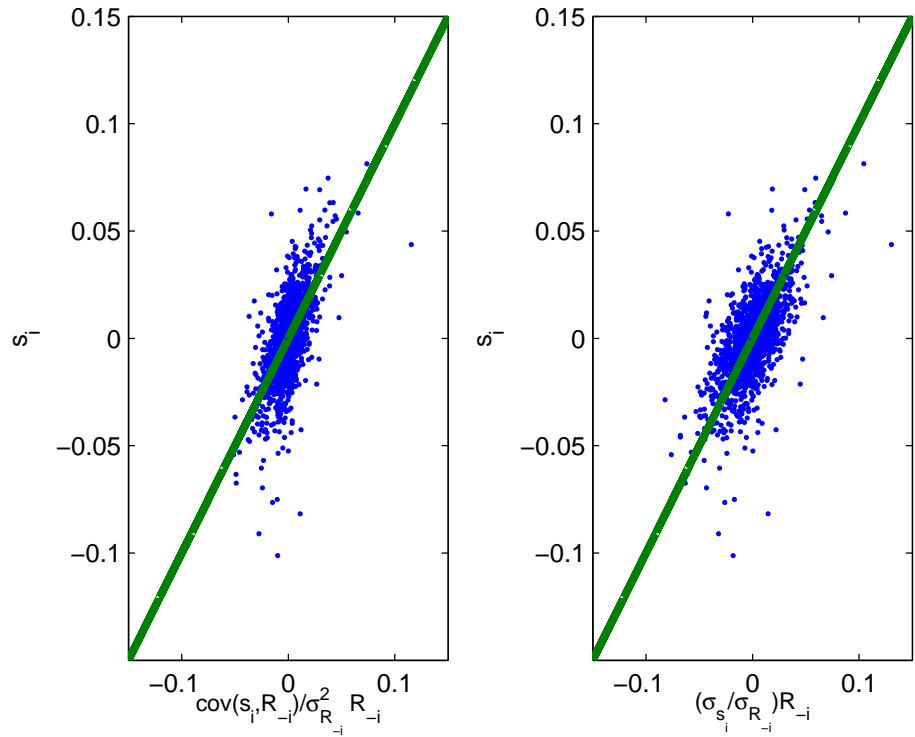}
\caption{\protect\label{Fig1}
The plot to the left shows the CAPM hypothesis (\ref{capm1}) using the daily returns 
of the Dow Jones Industrial index over the period 
03/01/2000 to 20/06/2008. 
Plot to the right illustrates 
instead (\ref{our_hyp}) using same data set. Each point correspond to a daily return 
$s_i$ of a given stock $i$. 
}
\end{figure}

\begin{figure}[h]
\includegraphics[width=14cm]{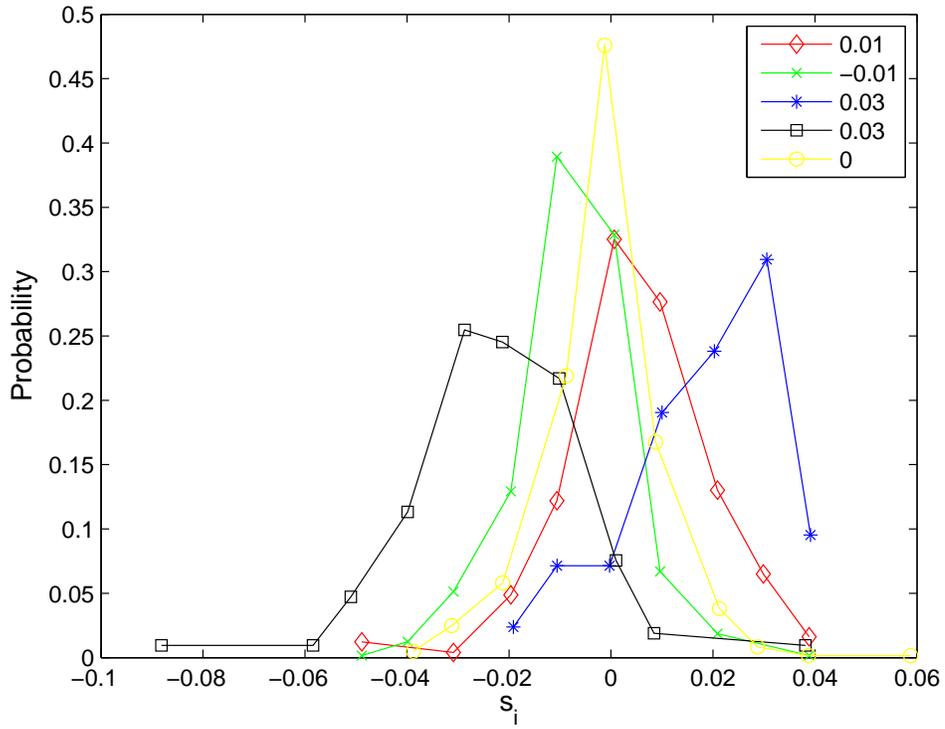}
\caption{\protect\label{Fig1_inset}
Probability distribution function of $s_i$ conditioned on 5 different values of 
$x \equiv {R_{-i} \over \sigma(R_{-i}) } 
 \sigma(s_{-i}) $
This figure is inset to figure 1.
}
\end{figure}

The data points in figure~\ref{Fig1} show the 
CAPM hypothesis (\ref{capm1}), and our hypothesis 
(\ref{our_hyp}) respectively, using daily returns of the 30 stocks of 
the Dow Jones Industrial Average over 
almost a decade of data\cite{note3}. A perfect fit of the data to the two equations would 
in both cases lie on the green diagonal. The data for CAPM appear tilted with respect to 
the diagonal, whereas the data concerning our hypothesis appear to be symmetrically 
distributed around the 
diagonal, which is what one should expect if the data on average followed (\ref{our_hyp}). 
Figure~\ref{Fig1}b therefore  
 gives some first evidence for the  support of (\ref{our_hyp}). This  
impression is strengthened when one takes a closer look at the cloud of data points in 
figure~\ref{Fig1}b, and consider the probability
 that the return of stock $i$ takes the value $s_i$,  
{\em conditioned on} a given fixed value of 
$x \equiv {R_{-i} 
 \over \sigma(R_{-i})  }
 \sigma (s_i) $. 
Figure~\ref{Fig1_inset} shows the probability distribution function of $s_i$ 
conditioned on  
five different values of the variable $x$. 
From the hypothesis (\ref{s_3})-(\ref{our_hyp}) one would expect the most likely value of the 
stock return $s_i$ to occur for the given fixed value of $x$. This is indeed seen 
to be the case with all five distributions peaking, if not actually at then close to $x$, 
giving further evidence to the assumption (\ref{our_hyp}).   

\begin{figure}[h]
\includegraphics[width=14cm]{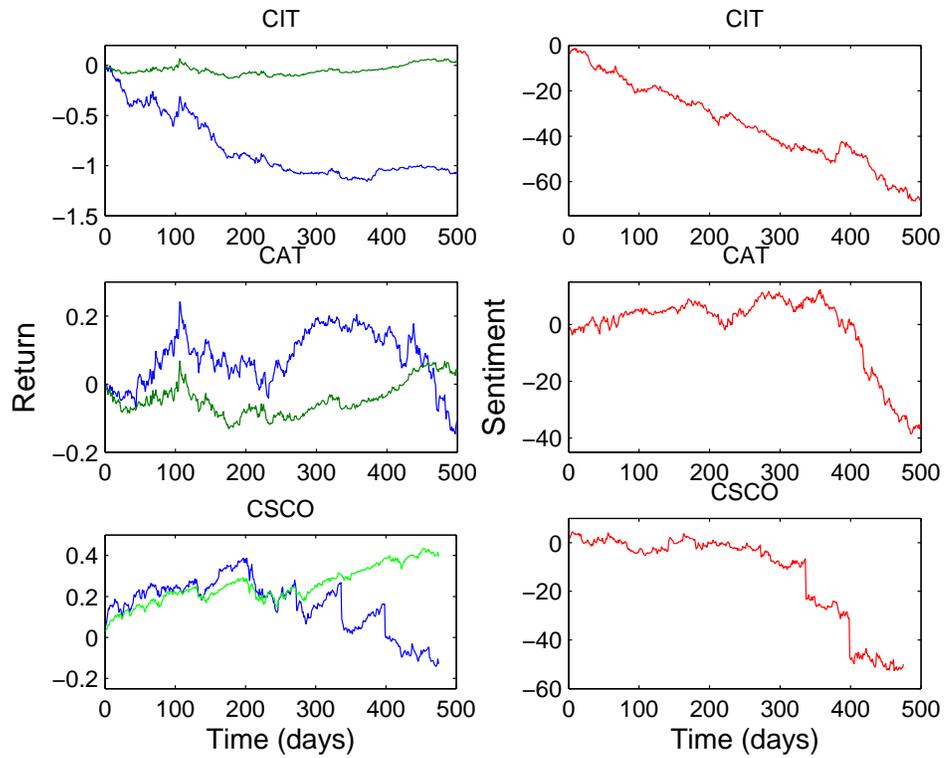}
\caption{\protect\label{Fig2}
The figure shows the cumulative sentiment for three different stocks, Citi Bank, 
 Caterpilar 
 in 
the time period (03/01/2000 to 20/06/2008) and Cisco 
 in 
the time period (01/06/2009 to 02/06/2011). The plots to the left shows in green 
the return of the Dow Jones and in blue the given stock over the given time period.  
}
\end{figure}

The sentiment $\alpha$ in (\ref{alpha}) was introduced as a behavioral trait, and as such we would 
expect to see its effect on a long term time scale say at least of the 
order weeks or months. 
Figure~\ref{Fig2} shows the cumulative sentiment for three different stocks, Citi Bank, 
and Caterpilar 
 in 
the time period (03/01/2000 to 20/06/2008) and Cisco 
 in 
the time period (01/06/2009 to 02/06/2011). The plots to the left shows in green 
the return of the Dow Jones and in blue the given stock over the given time period.  
The case of the Citi Bank stock is particular striking with a constant negative sentiment seen  
by the continuous decline in the cumulative sentiment curve of figure~\ref{Fig2}, corresponding 
to a constant sub-performance over two years. It should be noted that the data was chosen 
in order to have both a declining general market, which happens over the first half of the 
time period shown, as well as an increasing market which happens over 
the rest of the time period chosen.  It is  remarkable that the negative 
 sentiment of the Citi Bank 
stocks stays constant independent of whether the general trend is bullish or bearish. Similarly 
it should be noted that the general sentiment of Caterpilar  
had a neutral value in the declining market, but then developed a 
distinguishable negative sentiment over the last three or four months of the time series where 
the general market is bullish. The price history for Cisco Systems tells a similar story. The only 
difference here is the two big jumps happening after 350 and 400 days respectively. These 
two particular events took place on the 11/11/2010 and on the 10/2/2011. On the 11/11/2010 
the price dropped because of a bad report for the third quarter earnings. This gave 
rise to a loss of confidence by investors who  
 were expecting a sign of recovery after a couple of hard 
months. On the 10/2/2011  
Cisco Systems announced a drop in their earnings (down 18\%) together with a downward 
revision (-7 \%) of 
sales for their core product. It is worth noticing that the decline of the cumulative  
sentiment took place {\bf before} the two events: prior to 11/11/2010 there was a long 
 slow descent of the cumulative sentiment (meaning a constant negative sentiment) 
and after the 11/11/2010 the descent continued. 
This could be taken as evidence that some investors with insider knowledge were aware 
of the problems of the company, which was revealed only to the public on the 
two aforementioned 
 days.

\begin{figure}[h]
\includegraphics[width=14cm]{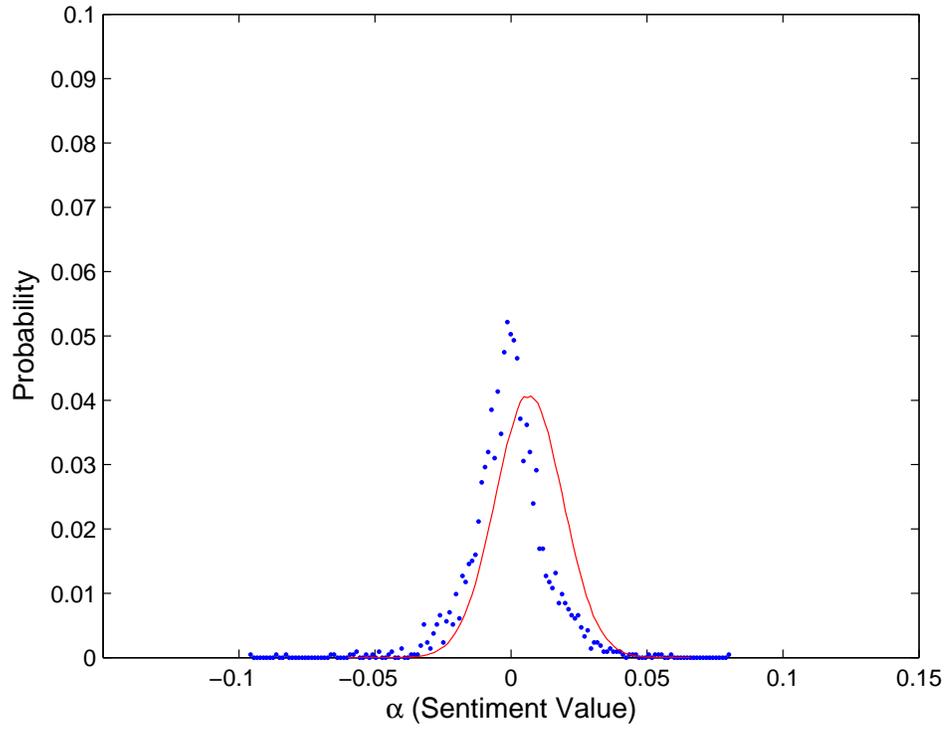}
\caption{\protect\label{Fig3}
Probability distribution function of the sentiment variable $\alpha_i$. 
}
\end{figure}

Figure \ref{Fig3} shows the probability distribution function of the sentiment variable 
obtained by sampling the sentiment variable $\alpha$ 
defined in (\ref{alpha}), 
using the daily return of all stocks of the Dow Jones Industrial Average in the 
period 03/01/2000-20/06/2008. 
As can be seen from the inset of figure~\ref{Fig3} the distribution appears  
to follow an exponential distribution for both positive and negative sentiments. One notes 
that the empirical distribution appears to be symmetric with respect to the sign of the 
sentiment - something which was implicitely assumed 
in deriving (\ref{our_hyp}) from (\ref{s_3}).

\section{Discussion}


The psychological experiments initiated by Kahneman and Tversky\cite{TK} more than 
three decades ago showed that not only do people not behave rationally, but what is even 
more important: people do not behave randomly - people are suceptible to common judgment errors.   
Therefore research on sentiments is important because neither the efficient market hypothesis 
nor the noise trader theory explains systematic deviations of asset prices from 
fundamental values. So far sentiments in financial markets have 
been explained as  either due to human overreaction or  
underreaction. For example Barberis et al.\cite{Barberis} used psychological reasoning 
in explaning market deviations: overreaction with representativeness heuristics and 
underreaction with conservatism. 

In this paper we have instead pointed out the importance of a 
{\em relative}  sentiment 
measure of a given stock to its peers. 
The idea is that people use heuristics, or ``rules of thumb'', in terms of ``yard sticks'' 
from the performance of the other stocks in a stock index. The under-/over-performance 
with respect to a yard stick then signifies a general negative/positive sentiment of 
the market participants towards a given stock. The bias created in such cases does 
not necessarily have a psychological origin but could be   
 due to insider information. Insiders having superior information about 
the state of a company reduce/increase their stock holding gradually 
 causing  a persisting bias over time. 
The introduction of a measure for the relative sentiment of a stock 
has allowed us to come up with a pricing 
formular for stocks very similar in structure to the CAPM model. 
Using empirical data of  
 the Dow Jones Industrial Average, stocks are shown to have daily performances 
with a clear tendency to cluster around the measures introduced by the yard sticks in 
accordance with our pricing formular.

\section{Acknowledgments}
J.V.A., M. M.  and S. M. would like to thank the Coll\'ege Interdisciplinaire de la Finance for financial 
support.

{}

\begin{figure}[h]
\includegraphics[width=14cm]{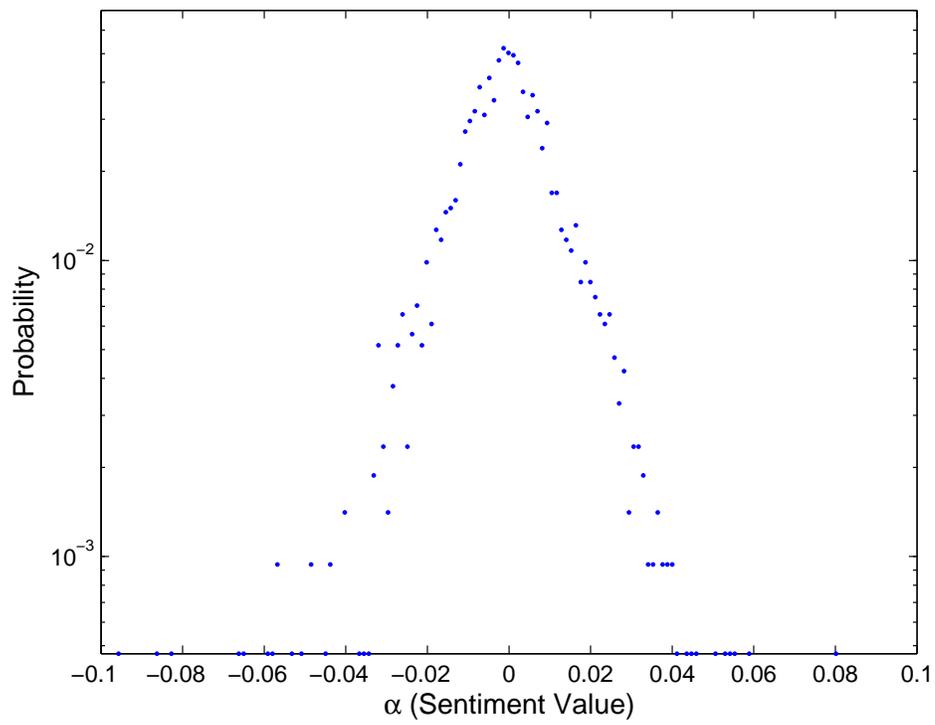}
\caption{\protect\label{Fig3_inset}
This figure is inset to figure 3. 
}
\end{figure}

\end{document}